\documentclass[a4paper]{spie}  

\usepackage{amsmath,amsfonts,amssymb}
\usepackage{graphicx}
\usepackage[colorlinks=true, allcolors=blue]{hyperref}

\title{Scientific Text Analysis with Robots\\ applied to observatory proposals}

\author[]{T. Jerabkova}
\author[]{H.M.J. Boffin}
\author[]{F. Patat}
\author[]{D. Dorigo}
\author[]{F. Sogni}
\author[]{F. Primas}

\affil[]{European Southern Observatory, Karl-Schwarzschild-str. 2, 85748 Garching, Germany}

\authorinfo{Please send correspondence to T.J., at Tereza.Jerabkova AT eso.org, or H.M.J.B., at hboffin AT eso.org.}

\begin{document} 
\maketitle

\begin{abstract}
To test the potential disruptive effect of Artificial Intelligence (AI) transformers (e.g., ChatGPT) and their
associated Large Language Models on the time allocation process, both in proposal
reviewing and grading, an experiment has been set-up at ESO for the P112 Call for Proposals.
The experiment aims at raising awareness in the ESO community and build valuable
knowledge by identifying what future steps ESO and other observatories might need to take
to stay up to date with current technologies. We present here the results of the
experiment, which may further be used to inform decision-makers regarding the use of AI in
the proposal review process. We find that the ChatGPT-adjusted proposals tend to receive lower grades compared to the original proposals. Moreover, ChatGPT 3.5 can generally not be trusted in providing correct scientific references, while the most recent version makes a better, but far from perfect, job. 
We also studied how ChatGPT deals with assessing proposals. It does an apparent remarkable job at providing a summary of ESO proposals, although it doesn't do so good to identify  weaknesses. When looking at how it evaluates proposals, however, it appears that ChatGPT systematically gives a higher mark than humans, and tends to prefer proposals written by itself. 
\end{abstract}

\keywords{Artificial intelligence, Large Language Model, proposal, observatory, funding}

\section{The rise of AI}
In the last decade or so, Artificial Intelligence (AI) has pervaded more and more our daily and work lives. Such explosion was driven by a combination of new methods, ever increasing specialised computing power, and consequent financial backing from the major computing companies. In astronomy also, AI is starting to have an impact, particularly in those areas where massive data sets exist, for example the classification of objects or of light curves, the measurement of photometric redshifts, the analysis of numerical simulations, or helping adaptive optics systems. A review of AI applications in astronomy can be found in Baron (2019) \cite{2019arXiv190407248B}, Djorgovski et al. (2022) \cite{2022arXiv221201493D}, and Smith \& Geach (2023) \cite{2023RSOS...1021454S}, but we also invite readers to look at some of the caveats or pitfalls in Ntampaka (2022) \cite{Ntampaka22}, Scott \& Frolop (2024)\cite{2024arXiv240319937S}, and Hogg \& Villar (2024) \cite{2024arXiv240518095H}.

At the European Southern Observatory (ESO), we have been following this rather closely, with several international workshops organised in the past years\footnote{eso.org/sci/meetings/2019/AIA2019.html, indico.cern.ch/event/881752, eso.org/sci/meetings/2022/SCIOPS2022.html} and one to be organised later this year\footnote{eso.org/sci/meetings/2024/esogpt24.html}. 
The transformative potential of AI may also come in different areas, through generative AI and large language models (LLMs) like the generative pre-trained transformers (GPT). These are poised to revolutionize the way scientists, and thus also astronomers, engage with data processing, proposal writing, and application evaluation. The widespread adoption of ChatGPT, released by OpenAI in late 2022, exemplifies the public's growing awareness of AI's capabilities, even if there is clearly also some hype and the adoption rate is not as large as generally advertised \cite{Fletcher24}. Still, ChatGPT’s ability -- and that of its more recent brethren like Google Gemini, Microsoft Copilot, or Claude-3 --  to generate creative content in seconds, from essays to scientific proposals, has sparked discussions within the scientific community regarding the utilisation of such tools in research workflows, leading to studies on its potential \cite{2024NatSR..14.4998S} and caution about its use, including the fact that sharing proposal information with generative AI technology via the open internet violates the confidentiality and integrity principles of many scientific organisations\footnote{see, e.g., https://new.nsf.gov/news/notice-to-the-research-community-on-ai}.  A search on the web tells us that there are now plenty of resources that claim that one can use ChatGPT to write a research proposal, including on the OpenAI page itself. It is thus of interest to look at how the use of ChatGPT could change the workflow of observing proposals. 

\section{A subjective look at ChatGPT and Gemini}
Whoever has tried any of these new AI assistants, based on transformers and LLMs, has certainly gone through two stages: first, a sense of wonder at how the ``machine'' is able to comprehend what we ask it, and apparently answers some intelligible sentences and paragraphs. But, then quickly, one also realises all the limitations of these tools, justifying why Google Gemini\footnote{For this paper, we only used the free version of Google Gemini} has a note that {\it ``[it] may display inaccurate info''}, while OpenAI’s ChatGPT state that  {\it ``[it] can makes mistakes.''} But if mistakes or inaccurate information are maybe not dramatic when searching the web for a new recipe for the best possible {\it Beef Wellington}\footnote{As ChatGPT suggested one of the authors.}, it could make a huge difference in getting or not your proposal approved. Thus, scientists should perhaps consider twice before using such tools. As Google Gemini mentions, {\it ``Scientific rigour is paramount, and relying solely on AI for content creation could introduce inaccuracies''} and {\it ``while ChatGPT can identify connections, it may struggle to capture the unique creativity that drives discovery.''} In addition, ChatGPT has a tendency to be very verbose, which is generally incompatible with the drastic space limitations of most research proposals.

Most importantly, one should not forget that these models will generally provide non-substantiated statements, and when asked to provide scientific references they may sometimes just make them up. In one case, we asked Gemini to provide a reference about the importance of binary stars in the formation and evolution of planetary nebulae. It correctly mentioned Boffin \& Jones (2019)\cite{2019ibfe.book.....B}, but with a completely wrong title. When prompted to provide the full reference, it gave a completely different -- and still wrong -- title, and a totally wrong reference. When asked why it did that, its excuse was that the original reference (a book) was not readily accessible, so it decided to provide a link to a paper to which it had access. It is doubtful that such an argument would convince a proposal reviewer. 

ChatGPT can in this regard perhaps be considered more damaging, at least ChatGPT 3.5 in this case. It seems that it creates references by randomly associating authors, date, titles, and journal references -- perhaps not so surprising for a statistical model of language. For example, if asked to provide one paragraph about the use of AI in proposal writing, this is what it provides: 
\begin{quote}
The integration of artificial intelligence (AI), exemplified by models like ChatGPT, into astronomical proposal writing and evaluation presents both opportunities and challenges. ChatGPT offers astronomers a tool to streamline proposal drafting, leveraging its capacity to synthesize astronomical data and provide feedback. However, concerns regarding AI-generated content quality, biases, and ethical considerations surrounding authorship emerge. Despite these challenges, the potential for ChatGPT to enhance proposal quality and foster collaboration within the astronomical community is significant (e.g., Medeiros, 2020; Seabrooke et al., 2021).

References: 
\begin{itemize}
	\item Medeiros, B. (2020). "AI in Astronomy: Artificial Intelligence Meets Astronomical Big Data." Annual Review of Astronomy and Astrophysics, 58, 31-60.
	\item Seabrooke, L., et al. (2021). "Machine Learning for Identifying Strong Gravitational Lenses in the DESI Legacy Imaging Surveys." Monthly Notices of the Royal Astronomical Society, 504(1), 1288–1298.
 \end{itemize}
\end{quote}

Apart from the fact that the statements are rather vague and unsubstantiated, the references mentioned do not exist at all, despite their apparent rigour. One can always hope that this will fool any reviewer, but we are certainly not far from intellectual dishonesty if we were to use this. It seems that ChatGPT 4o does a better job at not giving false references, as it gives links to actual papers. However, they are not necessarily up-to-date nor the most comprehensive. This could be damaging in a proposal, if the reviewer knows the field.

\section{An experiment}
ESO conducts regular calls for proposals to allocate observing time at the telescopes of its La Silla Paranal Observatory\footnote{eso.org/sci/observing/phase1/}. Given the growing interest in leveraging artificial intelligence tools, in particular Large Language Models, to assist in the proposal reviewing and grading process, we decided to investigate the impact of using an AI language model, specifically ChatGPT, on the quality of scientific rationales in proposals submitted to ESO. The objective is to determine whether AI can improve the clarity, appeal, and overall quality of proposals, thereby potentially influencing their success in the time allocation process.

To this end, a working group\footnote{Denominated STAR@ESO (STAR for ``Scientific Text Analysis with Robots'')} comprising the authors was formed to design and execute an experiment. This experiment  involved selecting test proposals from ESO Principal Investigators (PIs) and applying ChatGPT to try to ``enhance'' their scientific rationales. 

\setcounter{footnote}{0} 

\subsection{Methods}

The experiment was designed to assess the effectiveness of ChatGPT in improving the scientific rationales of proposals. The procedure consisted of the following steps:

1. {\bf Selection of Test Proposals}: Five proposals with ESO Principal Investigators were arbitrarily chosen for the experiment. The selected topics included: open star clusters, 
   stellar evolution, 
   exoplanets, 
   globular clusters, 
   and galaxies.

2. {\bf Defining Commands for ChatGPT}: A set of commands was formulated to instruct ChatGPT\footnote{In three cases, we used ChatGPT 3.5, and in two, we used ChatGPT 4.} on how to enhance the scientific rationales. Specific instructions included:
    \begin{itemize}
   \item Improving the text, asking the tool to make the text more appealing to reviewers, who are astronomers; 
   \item Finding and citing appropriate scientific references\footnote{As mentioned above, this generally meant introducing false references in the text. This was done in one case only.};
   \item Structuring the rationale with an introduction, detailed science goals, expected outcomes, and a final summary.
    \end{itemize}
    
3. {\bf Processing Proposals with ChatGPT}: The scientific rationales from these five ESO proposals were processed using ChatGPT. The adjusted rationales were then reviewed by us and compared with the original versions. The titles and abstracts of the proposals were also modified using ChatGPT.

4. {\bf Evaluation}: The adjusted rationales were evaluated based on their grades, which were assigned through the Distributed Peer Review\footnote{eso.org/sci/observing/phase1/distributed-peer-review.html} (DPR) process. The grades of the original and ChatGPT-adjusted proposals were compared to assess the impact of AI on the proposal quality.

\subsection{Results}

\subsubsection{Some notes on how ChatGPT changes proposals}
Although it is difficult to look in detail as how ChatGPT has changed all the proposals, it is interesting to take away some key messages.

{\bf Abstract.} Concerning the abstract, the AI tools have a tendency to be much more verbose, and one needs to ask it to be more concise. 
We have asked them, for example, to provide a better version of this abstract\footnote{Taken from a former ESO proposal from one of the authors -- it was not among the five proposals of the experiment described here.}, which has 101 words (669 characters):
\begin{quote}
    After the recent discovery of prominent spiral arms in the disc of the accreting binary IP Pegasi, we aim to study additional systems using Doppler imaging methods to resolve the detailed behaviour of the accretion disc and the impact region betweem\footnote{sic.} the disc and the mass transfer stream. Using high-resolution spectroscopy, we can use the dynamical information provided in the broad strong emission lines to spatially resolve the structure of the accretion flow. Several recent theoretical efforts on the properties of such spiral waves and on the disc-stream impact region provide various conclusions/predictions which can be directly tested using these observations.
\end{quote}

Here is the Gemini version, for comparison:
\begin{quote}
    The recent discovery of prominent spiral arms in the accreting binary IP Pegasi has opened a new window into the dynamics of accretion discs and their interaction with mass transfer streams. We propose high-resolution, optical spectroscopy with the XSHOOTER instrument on the ESO VLT to conduct Doppler imaging of these spiral arms.

This cutting-edge approach will allow us to spatially resolve the detailed behavior of the accretion disc in IP Pegasi, including the crucial impact region where the disc and the mass transfer stream collide. By analyzing the broad, strong emission lines in the spectra, we can extract crucial dynamical information and map the gas flows within the disc. This high-resolution data will enable us to directly test and differentiate between the various theoretical predictions on the properties of spiral waves and the disc-stream interaction region.

Our observations will not only provide groundbreaking insights into the specific case of IP Pegasi but also establish a powerful new technique for studying accretion disc dynamics in a wider range of binary systems.
\end{quote}

This version has 170 words and 1095 characters, a 70\% increase. Note that it made up the part about XSHOOTER based on a previous discussion we had with it (see Sec.~\ref{concl}), as this proposal was in fact to use UVES; XSHOOTER is not really a high-resolution spectrograph. More worrisome, is that the original abstract stated that we aim at studying {\it other systems than IP Pegasi}, while the newly worded abstract indicates that {\it we aim at studying IP Pegasi}. And the idea was not to study ``a wider range of binary systems'', but only ``additional systems''. A lot of caution is hereby warranted, even if some of the wording is likely useful to rewrite the original abstract. 

Let us now look at what ChatGPT 3.5 proposed\footnote{The outcome may be different if a more updated version is used, such as 4 or 4o, but as not everyone would have a paid version, this is likely useful.}:
\begin{quote}
    Recent discoveries have revealed striking spiral arms within the accretion disc of the binary system IP Pegasi. This breakthrough opens a new avenue for understanding the intricate dynamics of accretion flows. We propose to extend this exploration to additional systems, employing high-resolution Doppler imaging to capture the fine details of the accretion disc and the critical interaction zone between the disc and the mass transfer stream.

Our goal is to harness the power of high-resolution spectroscopy to decode the complex dynamical information embedded in the broad, strong emission lines characteristic of these systems. This approach will allow us to spatially resolve the structure of the accretion flow with unprecedented clarity.

Recent theoretical advancements have made significant strides in predicting the properties of spiral waves and the disc-stream impact region. However, these predictions vary and require empirical validation. Our observations will provide the critical data needed to test these theoretical models, bridging the gap between theory and observation.

By securing telescope time, we aim to deliver transformative insights into the physics of accretion discs, offering a clearer understanding of the mechanisms driving these cosmic phenomena. The implications of our study extend beyond IP Pegasi, potentially informing models of accretion in a wide range of astrophysical contexts.
\end{quote}

We are now at 205 words and 1416 characters. A factor two increase compared to the original text, and this will no more fit in the abstract box of an ESO proposal. Although wordy, this rewriting seems more correct, even if the last sentence seems to come out of the blue and isn't fully accurate. It correctly highlights, however, that the original abstract was lacking a final sentence to better define the impact of the proposal. 

When complaining to ChatGPT that the abstract is too long and for a more concise version, here is what we get:
\begin{quote}
    Recent discoveries of prominent spiral arms in the accretion disc of IP Pegasi offer a unique opportunity to explore accretion dynamics. We propose to study additional systems using high-resolution Doppler imaging to resolve the detailed behavior of the accretion disc and the disc-stream impact region. By analyzing the dynamical information in broad emission lines, we aim to spatially map the accretion flow structure. Our observations will test recent theoretical models predicting spiral wave properties and disc-stream interactions, providing crucial empirical validation. This study promises to advance our understanding of accretion processes in binary systems.
\end{quote}

This is shorter indeed: 94 words and 669 characters, so as long as the original abstract. Likely would we have obtained the same result directly if we had specified the character limit, which highlights the importance of using the correct prompts. Thus, learning how to master such prompts may soon need to become another string of the bow of the modern astronomer. Here again, the new version closes with a useful sentence that shows the impact of the work. There are some liberties taken, however, as the original ``recent discovery'' became ``recent discover{\bf ies}''. We will see in the next section if such rewriting is appreciated by reviewers. 

{\bf Title:} It is also often remarkably easy to spot that a title was created by an AI tool, as they generally tend to follow the same rule. For example, asking to provide a title to the abstract they wrote above, 
here is what ChatGPT suggests:
\begin{quote}
    Mapping the Hidden Spiral Arms: High-Resolution Doppler Imaging of Accretion Dynamics in IP Pegasi
\end{quote}
and here are the suggestions by Gemini:
\begin{quote}
    - Unveiling the Secrets of Accretion: Doppler Imaging of Spiral Arms in IP Pegasi and Beyond

    - Dancing with the Stream: Resolving the Accretion Disc-Stream Impact in IP Pegasi and Similar Systems

    - Mapping the Accretion Flow: High-Resolution Spectroscopy of Spiral Arms in IP Pegasi.
\end{quote}

A pattern is quite obvious. As a rule, it may be worth considering removing the part before the colon, as it doesn't bring anything, nor is it often correct\footnote{For example, the spiral arms aren't hidden. And why would we want to dance with the stream?}. It is also sometimes unclear what is meant, for example in the first example from Gemini, what does ``and beyond'' mean? Having a reviewer wondering about such questions is never good. Given that the abstract was wrongly rewritten, it is also not surprising that the new suggested titles are also wrong. For comparison, the original, informative, title, was:
\begin{quote}
    Spiral Structures in the Discs of Accreting Binaries.
\end{quote}


\begin{figure}[htbp]
    \centering
    \begin{tabular}{cc}
        \includegraphics[height=8cm]{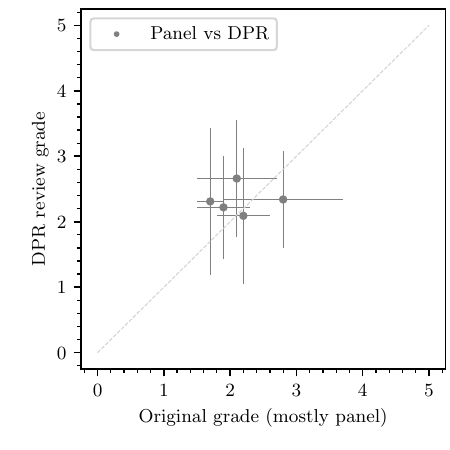} &
        \includegraphics[height=8cm]{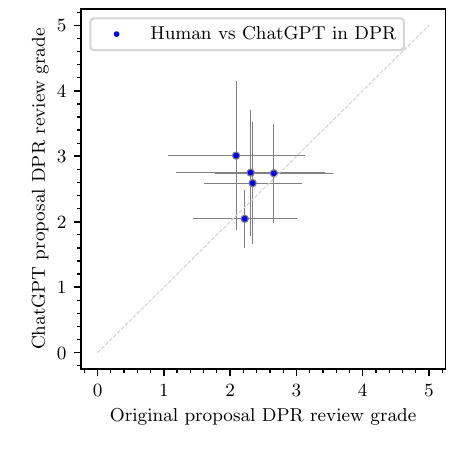} \\
    \end{tabular}
    \caption{
        \textbf{(a) Left Panel:} Scatter plot of the original grades of proposals (evaluated by a panel) against the DPR (Distributed Peer Review) grades for the same original (i.e., not ChatGPT-adjusted) proposals. \\
        \textbf{(b) Right Panel:} Scatter plot comparing the DPR grades of the original proposals against the DPR grades of the ChatGPT-adjusted proposals. A higher grade indicate a worst ranked proposal. In both panels, each point represents the average grade of a proposal, with error bars indicating the standard deviations. The gray dashed line represents the line of equality (x = y), indicating where the grades would lie if they were identical.
    }
    \label{fig:comparison}
\end{figure}

\subsubsection{Comparison of Original and ChatGPT-Adjusted Proposals}

The comparison of grades between the original proposals and the ChatGPT-adjusted proposals is illustrated in Figure \ref{fig:comparison}. The figure consists of two panels, one in which we compare the original grades of the proposal (which at the time were evaluated by a panel) to the grades obtained through the DPR, and another which compares the DPR grades of the original proposals and of the ChatGPT-adjusted proposals. 

\subsubsection{Bayesian Inference with MCMC}

To assess the significance of the differences in grades between the original and ChatGPT-adjusted proposals, we employed a Bayesian inference approach using Markov Chain Monte Carlo (MCMC) sampling. This method allows us to estimate the posterior distribution of the parameters of interest, specifically the mean difference in grades.

{\bf Model Specification: }
We modeled the differences in grades (\(\text{differences} = m_{\text{sai}} - m_{\text{s}}\)) using a normal distribution with an unknown mean (\(\mu\)) and standard deviation (\(\sigma\)). The likelihood function is given by:
\[
\text{Likelihood} = \mathcal{N}(\text{differences} \mid \mu, \sigma^2)
\]
where \(\mathcal{N}\) denotes the normal distribution.

{\bf Priors: }
We specified non-informative priors for the parameters:
\[
\mu \sim \mathcal{N}(0, 1)
\]
\[
\log(\sigma) \sim \mathcal{N}(0, 1)
\]
These priors reflect our initial uncertainty about the parameters.

{\bf MCMC Sampling: }
Using the \texttt{emcee} library\cite{2013PASP..125..306F}, we performed MCMC sampling to draw samples from the posterior distribution of the parameters. We initialized the walkers in a small Gaussian ball around the initial guess and ran the MCMC chain with 5000 steps, discarding the first 1000 steps as burn-in and thinning the chain by a factor of 10.

{\bf Posterior Analysis:}
The posterior distribution of the mean difference (\(\mu\)) was analyzed to determine the 95\% credible interval. The credible interval provides the range within which the true mean difference lies with 95\% probability. If this interval does not include zero, it indicates a statistically significant difference in grades.

\begin{figure}[ht]
    \centering
    \includegraphics[height=12cm]{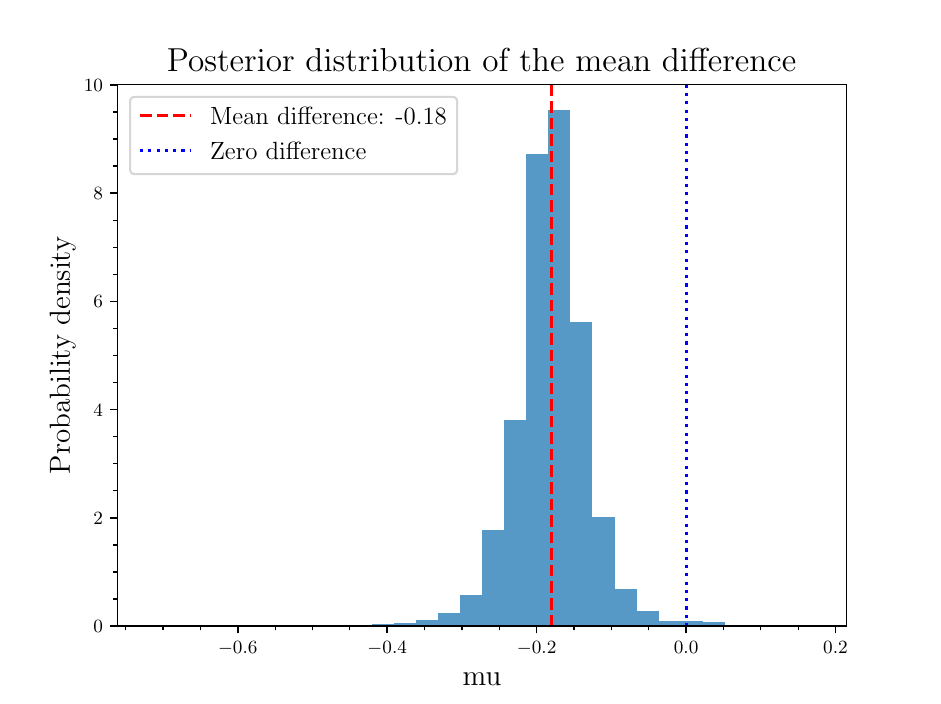}
    \caption{Posterior Distribution of the Mean Difference in Grades between Original and ChatGPT-Adjusted Proposals. This histogram shows the posterior distribution of the mean difference in grades between original and ChatGPT-adjusted proposals, estimated using MCMC sampling. The red dashed line indicates the mean difference, while the blue dotted line represents zero difference. The 95\% credible interval, which does not include zero, suggests that the difference in grades is statistically significant.}
    \label{fig:mcm_outcome}
\end{figure}

{\bf Results:}
The MCMC analysis revealed that the mean difference in grades between the original and ChatGPT-adjusted proposals is -0.18, with a 95\% credible interval of [-0.283, -0.076] (Fig.\ref{fig:mcm_outcome}). This interval does not include zero, suggesting that the difference is statistically significant and that the ChatGPT-adjusted proposals tend to receive lower grades compared to the original proposals.

This Bayesian MCMC approach provides a robust framework for assessing the significance of differences in grades, taking into account both the uncertainty and the variability in the data.

\section{Proposal assessment}
\begin{figure}[htbp]
    \centering
            \includegraphics[height=8cm]{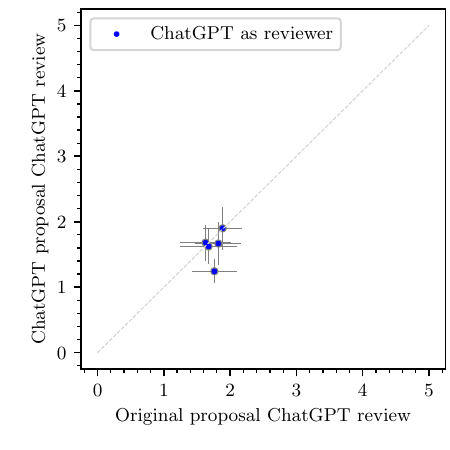} 
        \includegraphics[height=8cm]{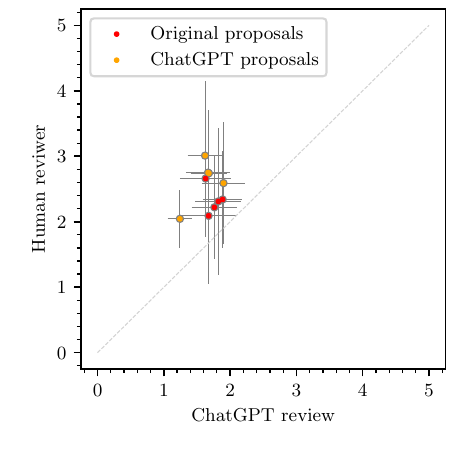} \\
       \caption{
        \textbf{Same as Fig.~\ref{fig:comparison}, but with the assessment done by ChatGPT itself. (a) Left Panel:} Grades by ChatGPT of the ChaGPT-adjusted proposals against the grades of the original proposals. \textbf{(b) Right Panel:} Scatter plot comparing the DPR grades (Human reviewer) of the original  (red) and ChatGPT-adjusted proposals (orange) against the ChatGPT grades of the same proposals. 
    }
    \label{fig:comparison2}
\end{figure}

In another experiment, we tried to see how ChatGPT can be used to assess proposals. 
The method is to ask ChatGPT to review the proposal as any referee who would like to ``cheat'' would do. It is important to stress that sharing proposals with a tool like ChatGPT is a breach of the confidentiality agreement that reviewers agree to follow. In this experiment, for each original and ChatGPT versions of the proposals mentioned above, we ask ChatGPT, separately to mimic different users, for 10 reviews\footnote{At ESO, in the DPR, each PI of a proposal is asked to review and grade ten other proposals -- see Jerabkova et al.\cite{2023Msngr.190...63J}.} with the following instructions (basically those that astronomers receive when they have to do the DPR): 
\begin{quote}
Hello, I would like you to provide a review of an ESO proposal. It should have around 800-1000 characters. Focus mainly on science relevance, quality and impact. Provide a grade from 1 to 5, one being the best and 5 being the worst, according to the following scale:
\begin{itemize}
    \item 1.0 – outstanding: breakthrough science
 \item 1.5 – excellent: definitely above average
 \item 2.0 – very good: no significant weaknesses
 \item 2.5 – good: minor deficiences do not detract from strong scientific case
 \item 3.0 – fair: good scientific case, but with definite weaknesses
 \item 3.5 – rather weak: limited science return prospects
 \item 4.0 – weak: little scientific value and/or questionable scientific strategy
 \item 4.5 – very weak: deficiences outweight strengths
 \item 5.0 – unsuitable. 
\end{itemize}
Provide a standalone review that can be passed to the proposal's author and then grade the evaluation for me to pass it to the ESO system. 
\end{quote}
We then copy and paste the title, abstract and science rationale in ChatGPT.
The results are quite interesting and illustrated in Figs.~\ref{fig:comparison2} and \ref{fig:mcm_outcome2}. 

\subsection{Results}
\begin{figure}[ht]
    \centering
    \includegraphics[height=7cm]{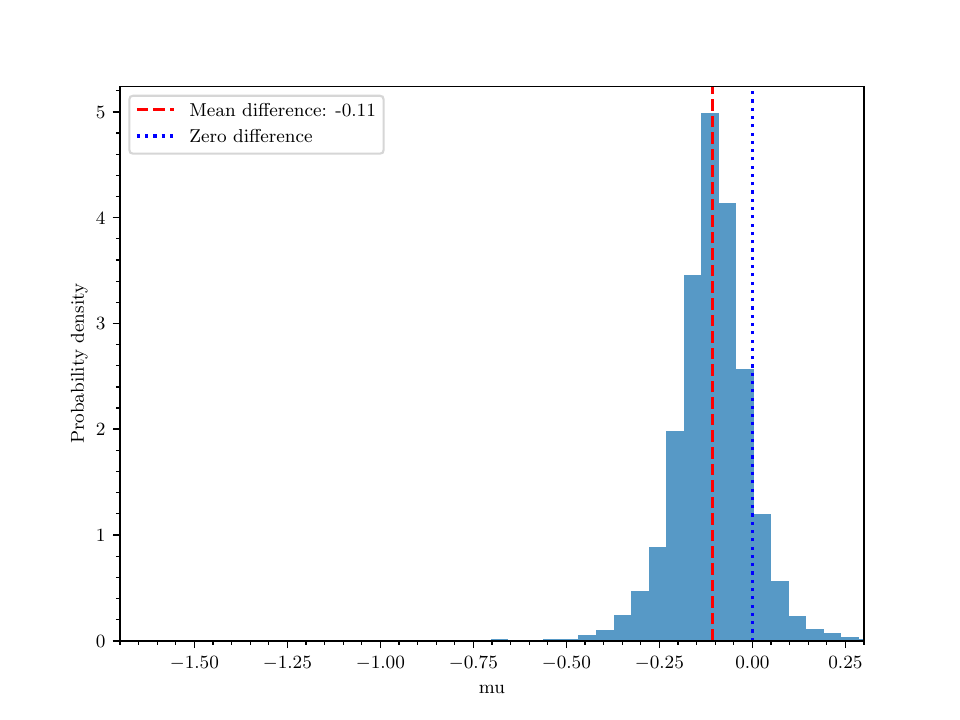}\\
     \includegraphics[height=6.3cm]{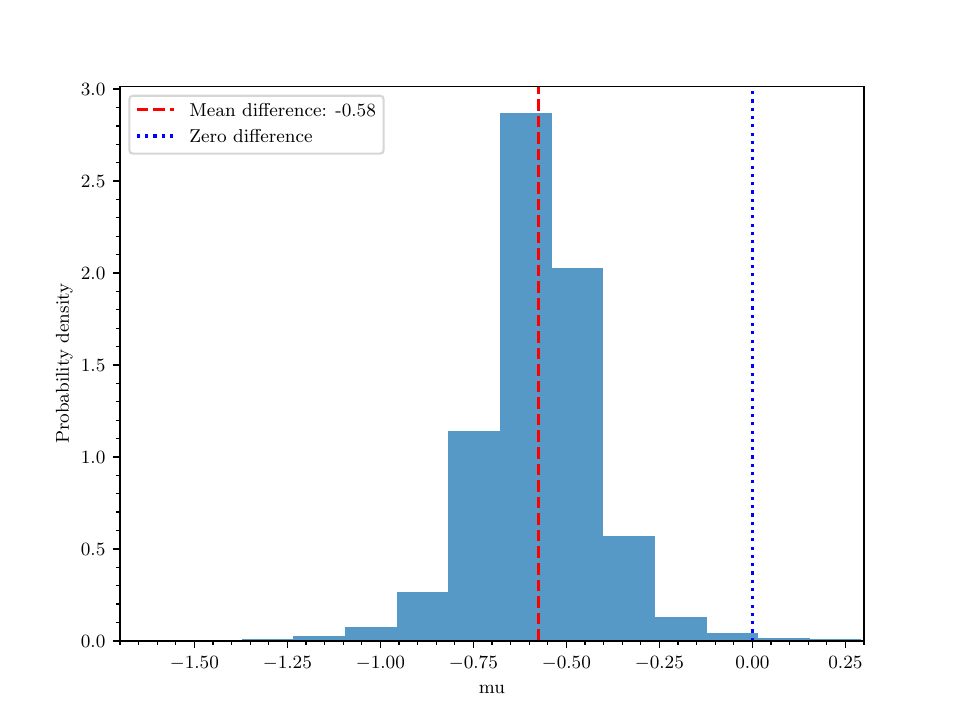}
      \includegraphics[height=6.3cm]{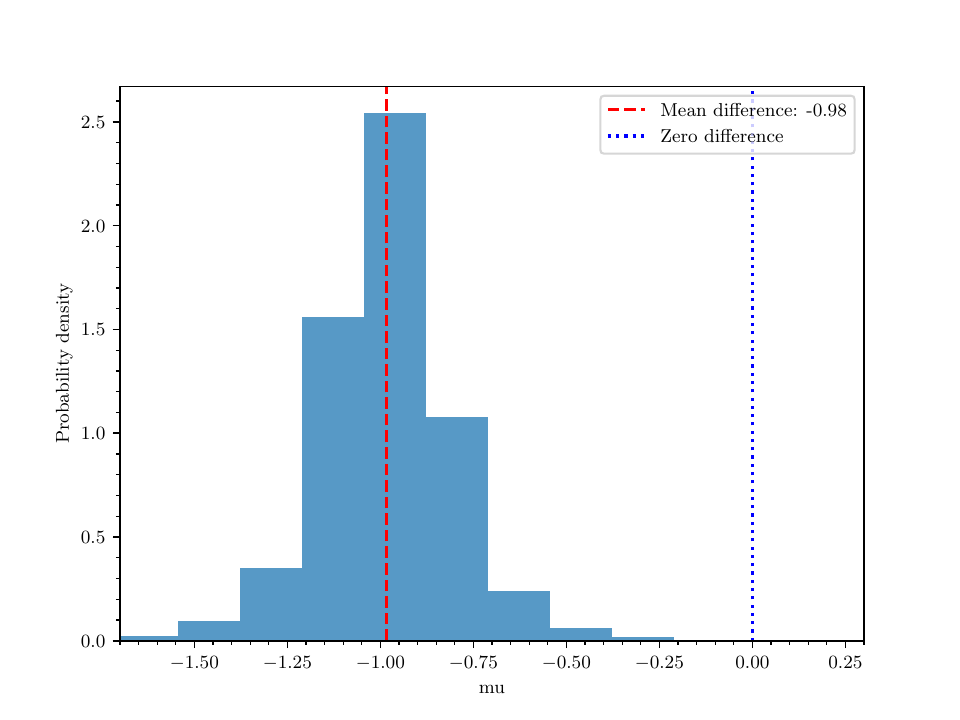}
    \caption{Similar figure to Fig.~\ref{fig:mcm_outcome}, but with the assessment done by ChatGPT itself. (top) Posterior Distribution of the Mean Difference between the ChatGPT reviews of  the original proposals and the ChatGPT-adjusted proposals. (bottom left) Same but showing the distribution of the mean difference between the DPR and ChatGPT grades of the original proposals. (bottom right) Same as the left panel but for the  ChatGPT-adjusted proposals. 
    }
    \label{fig:mcm_outcome2}
\end{figure}

Figure~\ref{fig:comparison2} clearly indicates that ChatGPT reviewed the original and ChatGPT-adjusted proposals almost the same (with no statistically significant difference as indicated in the top panel of Fig.~\ref{fig:mcm_outcome2}), although there is a slight indication that it would favor ChatGPT-adjusted proposals. Thus, although humans may not seem to see any improvements in the proposals, the tool seems to think it did a good job!

When looking at the grading consistency between GPT models and human reviewers, we observe that the GPT models, including ChatGPT versions 3.5, 4, and 4o, generate highly coherent reviews with greater consistency compared to human reviewers. Notably, version 3.5 tends to assign lower grades (higher values) on average compared to its more advanced counterparts, versions 4 and 4o.
However, it is important to highlight that ChatGPT systematically assigns higher grades to proposals than human reviewers. A critical test to further evaluate ChatGPT's performance would involve assessing its ability to identify errors and significant omissions in objectively poor proposals. This test would help determine its effectiveness in critical review scenarios.

The associated MCMC study is shown in Fig.~\ref{fig:mcm_outcome2}. It can be see the 95\% credible interval for the mean difference between the ChatGPT grades of both types of proposals is [-0.33,0.12], with a mean of -0.11, and the difference is thus not statistically significant. This is not the case when comparing the distribution of the mean difference between the DPR (i.e., humans) and ChatGPT grades for the original (95\% credible interval = [-0.91,-0.27]; with a mean of -0.58) and ChatGPT adjusted (95\% credible interval = [-1.38,-0.63]; with a mean of -0.99) proposals. The difference highlights the combination of two effects: ChatGPT prefers its own proposals, while humans penalise the latter, and the difference becomes thus greater. It would thus be dangerous to be in a world where an AI judges its own work. 

As an illustration, we provide here the outcome of the assessement of one proposal, using two different versions of ChatGPT:

{\bf ChatGPT-3.5}
\begin{quote}
Review:
The proposal aims to utilize UVES to confirm the presence of certain structures in a specific stellar system, employing a methodology previously tested via theoretical simulations. The scientific rationale is well-grounded, emphasizing the importance of such stellar systems in studying stellar evolution and galactic dynamics. The detection of unique structures around such systems provides valuable insights into their formation and dissolution processes, contributing to our understanding of galactic evolution.
The proposal's strength lies in its innovative approach, combining observational data with theoretical simulations to identify and study these structures. By targeting a subset of stars within the system, the authors plan to derive radial velocities and stellar abundances to confirm their membership in these structures. This confirmation would not only validate their methodology but also extend its applicability to other similar systems.
However, some potential limitations need addressing. While the proposal acknowledges the role of certain observational data in detecting these structures, it could provide more detailed discussion on how data limitations affect the identification and confirmation of candidate stars. Additionally, the proposal would benefit from a clearer explanation of how the selected subset of stars was chosen and how potential binary systems will be accounted for in the analysis.
Overall, the proposal presents a compelling scientific case with the potential for significant contributions to the field of stellar and galactic evolution. With careful consideration of the outlined limitations and further refinement of the experimental approach, this research has the potential for high impact.

Grade: 2.3 (Very Good: No significant weaknesses)
 \end{quote}
 
 {\bf ChatGPT-4o}
\begin{quote}
Science Relevance: The proposed study addresses a critical aspect of galactic astronomy: the dynamics and dissolution of stellar clusters within the Milky Way. The confirmation of specific stellar structures using advanced spectroscopic techniques is of high scientific relevance, as it provides insights into the processes governing cluster evolution and the interaction between clusters and the Galactic potential. This research has the potential to significantly enhance our understanding of stellar evolution, cluster dynamics, and the structure of the Milky Way.

Quality: The proposal is well-structured and builds on solid preliminary work, including simulations and data from high-precision astrometric surveys. The choice of the spectrograph for deriving radial velocities and stellar abundances is appropriate, and the methodology appears sound. The sample selection of 38 stars ensures a comprehensive analysis. The proposal addresses the limitations of existing data for faint stars and proposes a feasible observational strategy with a signal-to-noise ratio of 50 for accurate measurements.

Impact: Confirming the proposed stellar structures would provide crucial constraints on cluster dissolution rates and the initial mass function. The validation of the methodology and simulations can be extended to other clusters, offering a broader impact on the field. This work has the potential to advance our knowledge of the physical processes in the interstellar medium and the Galactic potential, contributing to a deeper understanding of the Milky Way’s structure and evolution.

Grade Evaluation: The proposal is robust, with a clear scientific case and a well-defined methodology. The potential impact on the field of galactic astronomy is significant, and the use of advanced spectroscopic techniques for this study is well-justified. While there are minor uncertainties regarding the confirmation of some stars as cluster members due to the possible presence of binaries, the overall quality and relevance of the proposal are outstanding.

Grade: 1.2

This grade reflects the proposal's high scientific value, solid methodological approach, and potential for significant impact on our understanding of cluster dynamics and galactic structure.
\end{quote}

As already mentioned earlier, both versions of ChatGPT came up with a very different grade. During the scheduling, the second review would mean that the proposal would end among the highest ranked ones, while the first review would likely signify that the programme would only be ranked in the second tier of all proposals, and would thus be scheduled, but only with lower priority. The improvement in the structure of the assessment between versions 3.5 and 4o is also remarkable. 

\section{Conclusion}\label{concl}
In this work, we have used ChatGPT in its various versions to assess whether it can be used to improve astronomical proposals or to assess such proposals as a distributed peer reviewer. 
We should stress that we only considered changing the title, abstract, and the scientific rationale, and did not consider the tools for choosing the instrument or the best observational strategy. Indeed, we should not expect a LLM to understand the scientific process, although we must admit that when, for example, we asked the distinction between the FORS2 and XSHOOTER instruments on ESO's Very Large Telescope, ChatGPT 3.5 didn't do too bad a job. Even though it still proposed that FORS2 offered a high-time resolution mode; a mode that has been decommissioned since 2018. More annoyingly, it recommends to use XSHOOTER over FORS2 for faint targets, which is clearly false. Apart from this, it was in a sense better than Google Gemini that claimed that the maximum resolution of XSHOOTER was 55,000 -- it is 18,000 -- and that FORS2 has a limited set of grisms, while in fact it has quite an extended range of available grisms of various spectral resolution and covering the whole optical range. On the other hand, Gemini correctly mentioned that FORS2 is better suited for faint targets, and provided links to the instrument web pages as well as to their exposure time calculators. This, unsurprisingly, shows that one cannot have it all. 

Coming back to the possible enhancement of the proposals, 
we have highlighted that the AI tools need to be given very specific instructions with regards to the title and abstract. There is also a risk that the tools will provide a completely different interpretation of the abstract, and even if they don't fall into this pitfall, they will introduce some incorrect statements. Furthermore, 
our experiment indicates that the use of ChatGPT leads to a worse grade when assessed by humans. It is thus perhaps not a good idea to rely on ChatGPT and its brethren for writing proposals, but it could be used for getting inspiration for partial rewrite of the text. Moreover, ChatGPT 3.5 (or Gemini) will simply make up references, and so cannot be used in a proposal to such purpose. Clearly, using such tools requires one to use the most advanced, paid versions. 

It may be useful nevertheless to add a cautionary note here: the proposals we used were all written by experienced scientists, very familiar with ESO facilities. This may mean that the proposals were intrinsically good, hence making it more difficult to improve them. Things would be rather different for a first time, junior and inexperienced scientist, possibly writing with weaker English. In that case, an AI-enhanced proposal may read  better, but this still needs to be done with a very critical mindset.

Regarding assessments, ChatGPT does quite a good job at providing a written feedback on a proposal. However, it is important to highlight that ChatGPT systematically assigns higher grades to proposals than human reviewers and is therefore of no much use in grading proposals, as the scores are too close to each other. 
We want to stress here again that using ChatGPT or other AI tools for review purposes violates the ESO reviewers' confidentiality agreement -- this is likely also the case for many other funding agencies. Thus, this could only be done on a local (i.e., not public) LLM. 

\acknowledgments 
We would like to thank our ESO colleagues Giacomo Beccari, Elena Valenti, and Jiri Zak for allowing us the use of their original ESO proposals to conduct the experiment in this study. 

\bibliography{report} 

\begin{thebibliography}{10}

\bibitem{2019arXiv190407248B}
{Baron}, D., ``{Machine Learning in Astronomy: a practical overview},'' {\em
  arXiv e-prints} ,  arXiv:1904.07248 (Apr. 2019).

\bibitem{2022arXiv221201493D}
{Djorgovski}, S.~G., {Mahabal}, A.~A., {Graham}, M.~J., {Polsterer}, K., and
  {Krone-Martins}, A., ``{Applications of AI in Astronomy},'' {\em arXiv
  e-prints} ,  arXiv:2212.01493 (Dec. 2022).

\bibitem{2023RSOS...1021454S}
{Smith}, M.~J. and {Geach}, J.~E., ``{Astronomia ex machina: a history, primer
  and outlook on neural networks in astronomy},'' {\em Royal Society Open
  Science}~{\bf 10},  221454 (May 2023).

\bibitem{Ntampaka22}
{Ntampaka}, M., ``{Machine Learning in Astronomy: Cautionary Tales for the
  Community},'' {\em STScI Newsletter}~{\bf 39},  1 (2022).

\bibitem{2024arXiv240319937S}
{Scott}, D. and {Frolop}, A., ``{Deeper Learning in Astronomy},'' {\em arXiv
  e-prints} ,  arXiv:2403.19937 (Mar. 2024).

\bibitem{2024arXiv240518095H}
{Hogg}, D.~W. and {Villar}, S., ``{Is machine learning good or bad for the
  natural sciences?},'' {\em arXiv e-prints} ,  arXiv:2405.18095 (May 2024).

\bibitem{Fletcher24}
{Fletcher}, R. and {Nielsen}, K., ``{AI and the future of news},'' {\em Reuters
  Institute for the Study of Journalism}  (2024).

\bibitem{2024NatSR..14.4998S}
{Schulze Balhorn}, L., {Weber}, J.~M., {Buijsman}, S., {Hildebrandt}, J.~R.,
  {Ziefle}, M., and {Schweidtmann}, A.~M., ``{Empirical assessment of ChatGPT's
  answering capabilities in natural science and engineering},'' {\em Scientific
  Reports}~{\bf 14},  4998 (Feb. 2024).

\bibitem{2019ibfe.book.....B}
{Boffin}, H. M.~J. and {Jones}, D.,  [{\em {The Importance of Binaries in the
  Formation and Evolution of Planetary Nebulae}}{\nolinebreak\hspace{0.1em}]}
  (2019).

\bibitem{2013PASP..125..306F}
{Foreman-Mackey}, D., {Hogg}, D.~W., {Lang}, D., and {Goodman}, J., ``{emcee:
  The MCMC Hammer},'' {\em PASP}~{\bf 125},  306 (Mar. 2013).

\bibitem{2023Msngr.190...63J}
{Jerabkova}, T., {Patat}, F., {Primas}, F., {Dorigo}, D., {Sogni}, F.,
  {Astolfi}, L., {Bierwirth}, T., and {Pr{\"u}mm}, M., ``{The First Results of
  Distributed Peer Review at ESO Show Promising Outcomes},'' {\em The
  Messenger}~{\bf 190},  63--66 (Mar. 2023).

\end{thebibliography}
\bibliographystyle{spiebib} 

\end{document}